\documentclass[11pt]{article}
\usepackage{fullpage}
\usepackage{latexsym,pst-node,amsmath,amssymb}

\title{A decision procedure for unitary linear quantum cellular automata%
\thanks{Part of this research was supported by the ESPRIT 
	Working Group RAND2.}}

\author{Christoph D\"urr%
\thanks{Universit\'e Paris-Sud,
	LRI, b\^at 490,
	F-91405 Orsay Cedex,
{\tt 	durr@lri.fr}.
{Part of this research was supported by the ISI Foundation.}
}
	\and
Miklos Santha%
\thanks{CNRS, URA 410,
	Universit\'e Paris-Sud,
	LRI, b\^at 490,
	F-91405 Orsay Cedex,
{\tt 	santha@lri.fr}.
{Part of this work was done while the author visited the Centre
de Recerca Matem\`atica, Institut d'Estudis Catalans, Bellaterra,
Spain.}
}
}

\date{}

\newcommand{\Qflip}{\mbox{\scriptsize\textsc{Qflip}}}
\newcommand{\CA}{{{\cal C}_{\negmedspace A}}}
\newcommand{\transition}[2]{\langle #1 | #2 \rangle}

\newcommand{\affHull}[1]{[#1]}
\newcommand{\linHull}[1]{\langle #1\rangle}

 \newcommand{\Choose}[2]{{#1 \choose #2}}

\newcommand{\idom}{\mathop{\mathrm{idom}}}

\newenvironment{keywords}{\textbf{Key words.}}{}
\newenvironment{AMS}{\textbf{AMS subject classification.}}{}
\newtheorem{theorem}{Theorem}	\newtheorem{corollary}{Corollary}
\newtheorem{lemma}{Lemma}	
\newenvironment{proof}{\noindent {\bf Proof }}{\hfill$\Box$\medskip}

\allowdisplaybreaks

\begin{document}
\maketitle
\thispagestyle{empty}

\begin{abstract}
Linear quantum cellular automata were introduced recently as one of
the models of quantum computing.  A basic postulate of quantum
mechanics imposes a strong constraint on any quantum machine: it has
to be \emph{unitary}, that is its time evolution operator has to be a
unitary transformation.  In this paper we give an efficient algorithm
to decide if a linear quantum cellular automaton is unitary.  The
complexity of the algorithm is $O(n^{\frac{3r-1}{r+1}}) = O(n^3)$ in
the algebraic computational model if the automaton has a continuous
neighborhood of size $r$, where $n$ is the size of the input.
\end{abstract}



\section{Introduction}

The classical models of computation, such as Turing machines,
random access machines, circuits, or cellular automata are all
universal in the sense that they can simulate each other with only
polynomial overhead.  These models are based on classical physics,
whereas physicists believe that the universe is better described by
quantum mechanics.

Feynman \cite{Fey82,Fey86} pointed out first that there might be a
substantial gap between computational models based on classical
physics and those based on quantum mechanics.  The \emph{quantum Turing
machine} (QTM), the first model of quantum computation, was
introduced by Benioff \cite{Ben82a,Ben82b}.  Deutsch in \cite{Deu85}
described a universal simulator for QTMs with exponential overhead.
Bernstein and Vazirani \cite{BV97} were able to construct a universal
QTM with only polynomial overhead.

Other quantum computational models were also studied recently.
Deutsch \cite{Deu89} has defined the model of \emph{quantum circuits},
and later Yao \cite{Yao93} has shown that QTMs working in polynomial
time can be simulated by polynomial size quantum circuits.  Physicists
were also interested in \emph{quantum cellular automata}: Biafore
\cite{Bia94} considered the problem of synchronization, Margolus
\cite{Mar94} described space-periodic quantum cellular automata and
Lloyd \cite{Llo93,Llo94} discussed the possibility to realize a
special type of quantum cellular automaton.  \emph{Linear quantum
cellular automata} (LQCAs) were formally defined by Watrous
\cite{Wat95} and by D\"urr, L\^eThanh and Santha \cite{DLS97}.  In the
former paper it was shown that a subclass of LQCAs, \emph{partitioned
linear quantum cellular automata} (PLQCAs) can be simulated by QTMs
with linear slowdown.  Van Dam~\cite{Dam96} defined space-periodic
LQCAs and gave a universal instance of this model.

We should make clear at this point that most of these models are only
theoretically motivated.  \emph{Real life} quantum computers as built
today in laboratories are essentially small quantum circuits or
partitioned cellular automata.

A quantum computational device is at any moment of its computation in
a \emph{superposition of configurations}, where each configuration
has an associated complex amplitude.  A superposition is \emph{valid}
if it has unit norm.  If the device is \emph{observed} then a
configuration will be chosen at random, where the probability of a
configuration to be chosen is equal to the squared magnitude of its
amplitude.  Therefore it is essential that valid superpositions be
transformed into valid superpositions, or equivalently, that the {\em
time evolution operator} of the device preserve the norm.  This
property is called \emph{well-formedness}, and thus it is a
natural problem to decide if a given quantum machine is well-formed.
In the case of QTMs and PLQCAs there exist easily checkable
constraints on the finite local transition function of the machine
which are equivalent to its well-formedness.  Such constraints were
identified respectively by Bernstein and Vazirani \cite{BV97} and by
Watrous \cite{Wat95}.  In the case of LQCAs no such local constraints
are known, still D\"urr, L\^eThanh and Santha \cite{DLS97} gave a
polynomial time algorithm to decide if an LQCA is well-formed.  Part of
this algorithm was improved by H{\o}yer \cite{Hoy96} using a different
approach.

However, one of the basic postulates of quantum mechanics imposes an
even stronger constraint than norm-preserving on the time evolution
operator.  It actually requires that this operator --- as any other
quantum operator --- be a unitary transformation.  We will call a
machine which satisfies this constraint \emph{unitary}.  In
\cite{BV97} and \cite{Wat95} it was proven that norm-preserving
already implies unitarity in the case of QTMs and PLQCAs.  It is also
trivially true for machines with finite configuration sets, such as
quantum circuits.  But this is not true for LQCAs; it is quite simple
to construct a well-formed LQCA which is not unitary.\footnote{A
well-known example is the classical LCA \textsc{Xor}$=(\{0,1\}, 0,
(0,1), \delta)$, where $\delta(x,y) = (x+y) \bmod 2$.  This cellular
automaton is injective for finite configurations but not surjective,
thus the associated LQCA is well-formed but not unitary.}

In this paper we give an efficient algorithm to decide if an LQCA is
unitary.  The complexity of our algorithm is cubic if the input LQCA
has continuous neighborhood (most papers in the literature about
classical linear cellular automata deal only with such cases).  Our
algorithm will use the procedure of \cite{DLS97} which in quadratic
time decides if the LQCA is well-formed.  The present paper actually
gives an algorithm which decides if a well-formed LQCA is also
unitary.

Well-formedness is equivalent to the orthonormality of the column
vectors of the time evolution operator; unitarity requires
orthonormality also from the row vectors.  Deciding unitarity is much
harder than deciding well-formedness.  One way of seeing this is that
whereas the column vectors have finite support, the row vectors can
have an infinite number of non zero components.

\section{The computational model}		\label{sec-model}

Let us fix for the paper the following notation.  If $u$ and $v$ are
vectors in some inner-product space over the complex or the real
numbers, then $\langle u| v\rangle$ will denote the inner product of
$u$ and $v$, and $\|u\|$ the norm of $u$.  If $M$ is a matrix in such a
space, then $M^*$ denotes its conjugate transpose.

We recall here the definition of a \emph{linear quantum cellular
automaton} (LQCA) which is the quantum generalization of the
classical one-dimensional cellular automaton.  A more detailed
description of this model can be found in \cite{Wat95} and in
\cite{DLS97}.

An LQCA is a 4-tuple $A=(\Sigma, q, N, \delta)$.  The \emph{cells} of
the automaton are organized in a line, and are indexed by the elements
of $\mathbb Z$.  The finite, non-empty set $\Sigma$ is the set of {\em
(cell-)states}, and $q\in\Sigma$ is a distinguished {\em
quiescent} state.  The \emph{neighborhood} $N=(a_1, \ldots, a_r)$
is a strictly increasing sequence of integers, for some $r\geq 1$,
giving the addresses of the neighbors relative to each cell.  This
means that the neighbors of cell $i$ are indexed by $i+a_1, \ldots ,
i+a_r$.  An automaton is \emph{simple} if its neighborhood is an
interval of integers, that is $a_r = a_1 + r-1$.  In this paper we deal
only with simple automata, and we will only explain briefly in the
conclusion how our results apply to the general case.

The states of the cells are changing simultaneously at every time step
according to the \emph{local transition function}.  This is the
mapping $\delta : \Sigma^{|N|} \rightarrow \mathbb C^\Sigma$, which
satisfies that for every $(x_1, \ldots ,x_r) \in \Sigma^r$, there
exists $y \in \Sigma$ such that $ [\delta (x_1 \ldots x_r)](y) \neq
0$.  If at some time step the neighbors of a cell are in states $x_1,
\ldots ,x_r$ then at the next step the cell will change into state $y$
with amplitude $[\delta(x_1 \ldots x_r)](y)$ which is denoted by
$\transition{\delta(x_1 \ldots x_r)}{y}$.  The quiescent state $q$
satisfies that
\[ \transition{\delta(q^r)}{y} =
\left\{                         \begin{array}{ll}
1 & \mbox{if } y=q,\\
0 & \mbox{if } y\neq q.
			\end{array} \right.
\]
The set of \emph{configurations} is $\Sigma^{\mathbb Z}$, where for
every configuration $c$, and for every integer $i$, the state of the
cell indexed by $i$ is $c_i$.  A configuration $c$ is \emph{finite} if
its \emph{support} $\{ i : c_i\neq q\}$ is finite.  We are dealing
only with LQCAs which evolve on finite configurations.  We will denote
the set of finite configurations by $\CA$, and from now on we use the
word \emph{configuration} to mean a \emph{finite configuration}.  For
a configuration $c$, let $\idom(c)$ be the \emph{interval domain} of
$c$, which is the smallest integer interval containing the support of
$c$.  For the sake of definiteness, we define the empty interval as
$[0,-1]$ which is the interval domain of the everywhere quiescent
configuration.

The local transition function induces the \emph{time evolution
operator} which we write in matrix form
$U_A : \CA \times \CA \rightarrow \mathbb C$,
where $U_A(d,c)$ is the transition amplitude of changing configuration
$c$ to configuration $d$ in one step.  It is
defined by
\[
	U_A(d,c) = \prod_{i\in\mathbb Z} \transition{\delta(c_{i+N})}{ d_i},
\]
where $\delta(c_{i+N})$ is a short notation for
$\delta(c_{i+a_1},\ldots, c_{i+a_r})$.  This product is well-defined
since $c$ has finite support.

The automaton evolves on \emph{superpositions of configurations} which
are elements of the Hilbert space $\ell_2({\cal C}_A)$. If at some
time step the automaton is in the superposition $u
\in {\mathbb C}^{\CA}$, then at the next time step it will be in
the superposition $U_A u$.  Therefore $U_A$ is also an operator on
$\ell_2(\CA)$.  $A$ is \emph{well-formed} if $U_A$ is norm-preserving,
and we say that it is \emph{unitary} if $U_A$ is a unitary
transformation.

We will work in the algebraic computational model where complex
numbers take unit space, and arithmetic and logical operations take
unit time.  The description size of an automaton is clearly dominated
by the local transition table $\delta$.  Therefore we define the
\emph{size} of the automaton to be $n=|\Sigma^{r+1}|$.

For the rest of the paper we will fix a well-formed simple LQCA
$A=(\Sigma, q, N, \delta)$.  Without loss of generality we assume
$N=(0,1,\ldots, r-1)$.  Indeed let $A'=(\Sigma,q,N',\delta)$ be the
well-formed simple LQCA with the general neighborhood
$N'=(j,\ldots,j+r-1)$ for some integer $j$.  We claim that $A$ is
unitary if and only if $A'$ is unitary.  Let $A''=(\Sigma, q, (j),
\delta')$ be the \emph{shift cellular automaton}, where $\delta'$ is
the identity.  $A''$ is unitary since $\delta'$ is unitary.  Moreover
$U_A = U_{A''} U_{A'}$ which proves the claim.

\subsection{Example}

The figures in this paper will illustrate our algorithm with the
following LQCA:\\ \textsc{Qflip}$=(\{a,b\}, a, (0,1), \delta)$ with
$\transition{\delta(x,y)}{z}$ defined for all $x,y,z\in\{a,b\}$ by the
table:
\[
	\begin{array}{l|cc}
	xy\backslash z&a&b\\ \hline
	aa& 1&0 \\
	ba& 0&1 \\
	ab& 1/\sqrt2 & 1/\sqrt2	\\
	bb& 1/\sqrt2 & -1/\sqrt2	
	\end{array}
\]
Using the algorithm in~\cite{DLS97} it can be shown that
\textsc{Qflip} is well-formed, and in this article we show that its
evolution operator $U_{\Qflip}$ is even unitary.  This LQCA has an
interesting property.  For $n\geq 0$, let $c^n$ be the
configuration which is $b$ in all cells of index $i\in[-n,-1]$ and is
$a$ elsewhere.  Then we have for all $n\geq 1$, $U_{\Qflip}(c^1, c^n)
= (1/\sqrt2)^n$.  Thus an infinite number of configurations lead with
non-zero amplitude to the single configuration $c^1$.

For the all quiescent configuration $c^0$, we have $U_{\Qflip}(c^0,
c^0)=1$, thus there is a unique configuration leading to it.
Therefore when the LQCA \textsc{Qflip} runs backwards in time, every
cell with index $i\leq -1$ depends on cell $-1$.  From this we conclude
that there cannot be a LQCA with finite neighborhood whose evolution
operator is exactly $U_{\Qflip}^*$.

This makes the model of LQCA different from QTMs, since for every
well-formed QTM $M$, there exists a QTM which runs $M$ backward in
time with a constant time overhead.  It explains a bit why it seems
difficult to simulate any LQCA by a QTM.

\section{The main result}

The main result of the paper is the following theorem.

\begin{theorem}						\label{thm-main}
  There exists an algorithm which takes a simple LQCA as input, and
  decides in time $O(n^{\frac{3r-1}{r+1}}) = O(n^3)$ if it is unitary.
\end{theorem}

Since in \cite{DLS97} a $O(n^2)$ algorithm is given to decide if an
LQCA is well-formed, we will give only an algorithm which decides if a
well-formed LQCA is unitary.  The following lemma states that we only
have to verify that the rows of the time evolution operator are of
unit norm.
\begin{lemma} \label{lem-iso}
  Let $U \in {\mathbb C}^{\CA \times \CA}$ be a linear
operator.  If $U$ is norm-preserving then its rows have norm at most
$1$.  If all the rows are of unit norm then $U$ is unitary.
\end{lemma}

\begin{proof}
Let $c$ be a configuration, and $\bf c$ the superposition which has
amplitude $1$ for $c$ and $0$ elsewhere.  Then the norm of the row indexed
by $c$ in $U$ is $\| U^* {\bf c} \|$.  Since $U$ is norm-preserving
$\| U^* {\bf c} \| = \| U U^* {\bf c} \|$ and the projection of $U
U^*{\bf c}$ on ${\bf c}$ has norm
\[
		|\langle {\bf c} | U U^* {\bf c} \rangle|
	=	|\langle U^* {\bf c} | U^* {\bf c} \rangle|
	= 	\| U^* {\bf c} \|^2.
\]
But the projection of $UU^*{\bf c}$ on a unit vector has norm at most
$\| U^*{\bf c}\|$, and therefore $\| U^*{\bf c}\| \leq 1$.

For the second part of the lemma observe that the projection of $U U^*
{\bf c}$ on ${\bf c}$ has norm $1$.  Since $U$ is norm-preserving the
projection on any other basis vector ${\bf c'}$ must be $0$.  Thus
$\langle {\bf c'} | U U^* {\bf c}\rangle$ is $1$ if $c=c'$ and $0$
otherwise, or in other words $U U^* = I$, which concludes the proof.
\end{proof}

The outline of the proof is the following.  First we give a sequence
of reduction steps in section~\ref{sec-reduc} to a graph theoretical
problem and to another one from linear algebra.  We then give an
algorithm to solve the problem.  The different steps of the algorithm
are presented in sections~\ref{sec-border}, \ref{sec-coset}
and~\ref{sec-indep}.  The proof of the main theorem is then summarized
in section~\ref{sec-together}.

\section{The reduction}
\label{sec-reduc}

\begin{figure}[htb]
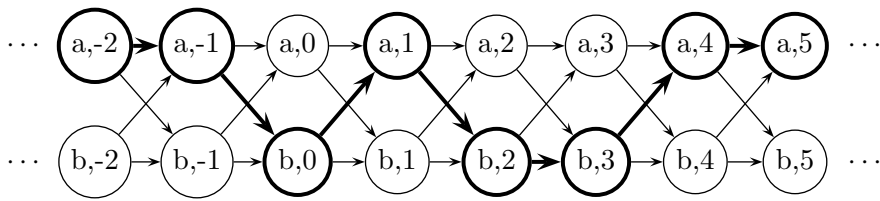

\begin{center}
\psset{linewidth=0.5pt}
\begin{tabular}{@{}c@{}cccccccc@{}c@{}}
$\cdots$&
\circlenode[linewidth=1.5pt]{a0}{a,-2}&	
\circlenode[linewidth=1.5pt]{a1}{a,-1}&	
\circlenode{a2}{a,0}&	
\circlenode[linewidth=1.5pt]{a3}{a,1}&	
\circlenode{a4}{a,2}&	
\circlenode{a5}{a,3}&	
\circlenode[linewidth=1.5pt]{a6}{a,4}&	
\circlenode[linewidth=1.5pt]{a7}{a,5}&$\cdots$\\[7mm]
$\cdots$&
\circlenode{b0}{b,-2}&	
\circlenode{b1}{b,-1}&	
\circlenode[linewidth=1.5pt]{b2}{b,0}&	
\circlenode{b3}{b,1}&	
\circlenode[linewidth=1.5pt]{b4}{b,2}&	
\circlenode[linewidth=1.5pt]{b5}{b,3}&	
\circlenode{b6}{b,4}&	
\circlenode{b7}{b,5}&$\cdots$\\	
\hspace*{9mm}&\hspace*{9mm}&\hspace*{9mm}&\hspace*{9mm}&\hspace*{9mm}&
\hspace*{9mm}&\hspace*{9mm}&\hspace*{9mm}&\hspace*{9mm}&\hspace*{9mm}
		    \ncline{->}{a0}{b1} 
\ncline{->}{a1}{a2} 
\ncline{->}{a2}{a3} \ncline{->}{a2}{b3} 
\ncline{->}{a3}{a4}  
\ncline{->}{a4}{a5} \ncline{->}{a4}{b5} 
\ncline{->}{a5}{a6} \ncline{->}{a5}{b6} 
\ncline{->}{a6}{b7} 
\ncline{->}{b0}{a1} \ncline{->}{b0}{b1} 
\ncline{->}{b1}{a2} \ncline{->}{b1}{b2} 
		    \ncline{->}{b2}{b3} 
\ncline{->}{b3}{a4} \ncline{->}{b3}{b4} 
		    \ncline{->}{b4}{a5} 
\ncline{->}{b5}{b6} 
\ncline{->}{b6}{a7} \ncline{->}{b6}{b7} 
\psset{linewidth=1.5pt}
\ncline{->}{a0}{a1} 
		    \ncline{->}{a1}{b2}
\ncline{->}{b2}{a3} 
		    \ncline{->}{a3}{b4}
\ncline{->}{b4}{b5}
		    \ncline{->}{b5}{a6} 
\ncline{->}{a6}{a7}
\end{tabular}
\end{center}
\caption{The configuration graph of the LQCA \textsc{Qflip}.  The
bold path corresponds to the configuration $\ldots aababbaa
\ldots$.}
\label{G_infty}
\end{figure}

\begin{figure}[htb]
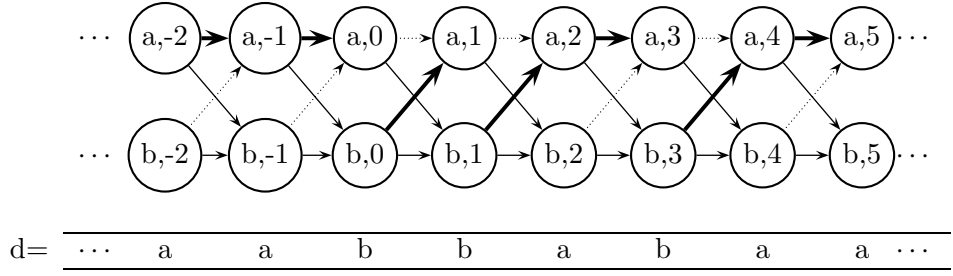

\begin{center}
\begin{tabular}{c@{}c@{}cccccccc@{}cc}
&
$\cdots$&
\circlenode{a0}{a,-2}&	
\circlenode{a1}{a,-1}&	
\circlenode{a2}{a,0}&	
\circlenode{a3}{a,1}&	
\circlenode{a4}{a,2}&	
\circlenode{a5}{a,3}&	
\circlenode{a6}{a,4}&	
\circlenode{a7}{a,5}&$\cdots$\\[7mm]
&
$\cdots$&
\circlenode{b0}{b,-2}&	
\circlenode{b1}{b,-1}&	
\circlenode{b2}{b,0}&	
\circlenode{b3}{b,1}&	
\circlenode{b4}{b,2}&	
\circlenode{b5}{b,3}&	
\circlenode{b6}{b,4}&	
\circlenode{b7}{b,5}&$\cdots$\\	
\hspace*{9mm}&\hspace*{9mm}&\hspace*{9mm}&\hspace*{9mm}&\hspace*{9mm}&
\hspace*{9mm}&\hspace*{9mm}&\hspace*{9mm}&\hspace*{9mm}&\hspace*{9mm}	\\
\cline{2-11}
d=&
$\cdots$&
a&a&b&b&a&b&a&a
&$\cdots$ \\ \cline{2-11}
\psset{linewidth=0.5pt}
\psset{doubleline=false}
\ncline{->}{a0}{b1} \ncline{->}{b0}{b1} 
\ncline{->}{a1}{b2} \ncline{->}{b1}{b2} 
\ncline{->}{a2}{b3} \ncline{->}{b2}{b3} 
\ncline{->}{a3}{b4} \ncline{->}{b3}{b4} 
\ncline{->}{a4}{b5} \ncline{->}{b4}{b5} 
\ncline{->}{a5}{b6} \ncline{->}{b5}{b6} 
\ncline{->}{a6}{b7} \ncline{->}{b6}{b7} 
\psset{linestyle=dotted,dotsep=1pt,linewidth=0.5pt}
\ncline{->}{b0}{a1}
\ncline{->}{b1}{a2}
\ncline{->}{a2}{a3}
\ncline{->}{a3}{a4}
\ncline{->}{b4}{a5}
\ncline{->}{a5}{a6}
\ncline{->}{b6}{a7}
\psset{linestyle=solid,linewidth=1.5pt}
\ncline{->}{a0}{a1} \ncline{->}{a1}{a2} 
\ncline{->}{a4}{a5} \ncline{->}{a6}{a7} 
\ncline{->}{b2}{a3} \ncline{->}{b3}{a4} 
\ncline{->}{b5}{a6} 
\end{tabular}
\end{center}
\caption{The configuration graph of the LQCA \textsc{Qflip} weighted
by the configuration $d=\ldots aabbabaa\ldots$.  The bold edges
have weight $1$, the normal edges have weight $1/2$ and the
dotted-line edges have weight $0$.}
\label{G_infty^d}
\end{figure}

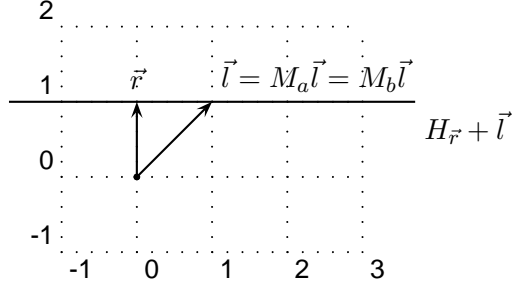
\begin{figure}[htb]
\begin{center}
\begin{pspicture}{}(-1,-1)(3,2)
\psgrid[griddots=5,subgriddiv=1](-1,-1)(3,2)
\psline{*->}(0,0)(0,1)	\uput[90](0,1){$\vec r$}
\psline{*->}(0,0)(1,1)	\uput[45](1,1){$\vec l = M_a \vec l = M_b \vec l$}
\psline{-}(-1.7,1)(3.7,1) 	\uput[315](3.7,1){$H_{\vec r}+\vec l$}
\end{pspicture}
\end{center}
\caption{For the LQCA \textsc{Qflip}, the border vectors $\vec l$ and
$\vec r$, and the affine subspace $H_{\vec r}+\vec l$.  In this
example $\vec l$ is a fix-point for the operators $M_a$ and $M_b$,
which shows that $U_{\Qflip}$ is unitary.}
\label{hyper}
\end{figure}

The different reduction steps are illustrated in the
figures~\ref{G_infty} to~\ref{hyper}.

Our problem is the following.  We have to decide if all row vectors of
the evolution operator associated to a given LQCA have unit norm,
under the assumption that the operator is norm preserving.  The naive
method fails because one would have to compute the norm for an
infinite number of rows.  Moreover for every row there can be an
infinite number of non-zero entries and every entry is defined by a
product on an unbounded number of terms.  The purpose of this section
is to reduce our problem to a finite one.

The \emph{configuration graph} is the infinite directed graph
$G_\infty(V,E)$ defined by $V= \Sigma^{r-1} \times \mathbb Z$ and
$E=\{((xt,i),(ty,i+1)) : x,y \in \Sigma, t\in \Sigma^{r-2} , i\in \mathbb
Z \}$.  To our knowledge this type of graph has been first used by
Sutner and Maas~\cite{SM88} to show that a particular robot motion
planning problem in the presence of moving obstacles is
\textsc{Pspace}-hard.  It was used again
by Sutner~\cite{Sut91} to prove that the predecessor of every
recursive configuration is also recursive.

A non-empty sequence (possibly infinite to the left, to the right or
in both directions) of vertices $(\ldots, (w_i, i), \ldots)$ in
$G_\infty$ is a \emph{path} if and only if for at at most a finite
number of indices $i$ we have $w_i \neq q^{r-1}$ and there is an edge
between every two immediate vertices.  Note that a sequence with a
single vertex is already a path.  We denote by $F,L,R$ and $P$
respectively the set of paths which are finite, infinite to the left,
infinite to the right and infinite to both directions.
Figure~\ref{G_infty} illustrates a path of $P$.

We say that two paths $p_1$ and $p_2$ are \emph{compatible} if the
last vertex of $p_1$ and the first vertex of $p_2$ exist and they are
the same.  In that case the \emph{composition} $p_1 \otimes p_2$ is
the concatenation of the two sequences after identifying the extreme
vertices.  If $P_1$ and $P_2$ are sets of paths, then
\[
P_1 \otimes P_2 = \left\{ p_1 \otimes p_2 :\begin{array}{ll} p_1 \in P_1,
p_2 \in P_2,\\ p_1
\mbox{ and } p_2 \mbox{ are  compatible}   \end{array}\right\}.
\]

Let $d$ be an arbitrary configuration.  It induces a \emph{weight}
function $g_d$ for the edges of $G_\infty$, where $g_d( (xt,i), (ty,
i+1) ) = |\transition{\delta(xty)}{d_i}|^2$.  We extend the weight function $g_d$
to paths and to sets of paths.  The weight of a path is the product of
the respective edge weights, and the weight of a path set is the sum
of the respective path weights.  The weight of a path consisting of a
single vertex is 1, and the weight of the empty path set is 0.  We
denote this \emph{weighted configuration graph} by $G_{\infty}^d$
which is illustrated in figure~\ref{G_infty^d}.

Although the weight of an infinite path is an infinite product, it is
well defined since all but a finite number of edges have weight 1.
The following lemma establishes a strong relationship between the
weight of an infinite path in $G_\infty^d$ and the entries of the time
evolution matrix.
\begin{lemma}					\label{lem-weightRow}
For any configuration $d$ the row indexed by $d$ in $U_A$ has norm
$\sqrt{g_d(P)}$.
\end{lemma}
\begin{proof}
We will show that there is a bijection $h$ between the set of
configurations ${\cal C}_A$ and the set of infinite paths $P$ in
$G_\infty^d$ such that for every configuration $c$ and $d$
\[
	g_d(h(c)) = | U_A(d,c) |^2.
\]
Summing up over all configurations $c$ will immediately conclude the
lemma.

Let $h : \CA  \rightarrow P$ be  defined for all  configuration
$c$ by
\[
	h(c) =  ( \ldots , (c_{i}\ldots c_{i+{r-2}}, i), \ldots ).
\]
Then it is a bijection, and the following equalities conclude the proof.
\begin{eqnarray*}
g_d(h(c))
&=& \prod_{i\in\mathbb Z}  
	g_d(  (c_{i}\ldots c_{i+r-2},i),
	      (c_{i+1}\ldots c_{i+r-1},i+1) )
 		\\
&=& \prod_{i\in\mathbb Z} |[ \delta(c_{i+N})] (d_i) |^2	\\
&=& \left| \prod_{i\in\mathbb Z}  
		[ \delta(c_{i+N})] (d_i)  \right|^2	\\
&=& |U_A(d,c)|^2.
\end{eqnarray*}
\end{proof} 

With Lemma~\ref{lem-iso} we got the following reduction of our
problem.
\begin{corollary}					\label{cor-path}
In a well-formed LQCA, for every configuration $d$, we have
$g_d(P)\leq 1$.  Moreover the automaton is unitary if and only if for
every $d$, $g_d(P)=1$.
\end{corollary}

Let us fix an interval $[j,k]$ and a configuration $d$ with
$\idom(d)\subseteq[j,k]$.  This interval induces a subset of the path
sets $L,F$ and $R$.  For every $w, w' \in \Sigma^{r-1}$ we set
\begin{eqnarray*}
L^j_w &=&  \{p\in L: \mbox{the last  vertex of }  p \mbox{ is }
(w,j) \},
	\\
F^{j,k}_{w,w'} &=& \left\{p\in  F: \begin{array}{@{}l@{}}
	\mbox{first  vertex of } p \mbox{ is } (w,j), \\
	\mbox{last vertex  of }  p \mbox{ is  } (w',k+1)
			\end{array} \right\},
	\\
R^k_{w'} &=& \{p\in R: \mbox{the first vertex of } p
\mbox{ is } (w',k+1) \}.
\end{eqnarray*}
Since the set of infinite paths can be decomposed as
\[
P = \bigcup_{w,w'\in\Sigma^{r-1}}
L^j_w \otimes F^{j,k}_{w,w'} \otimes R^k_{w'},
\]
we have
\[
g_d(P) = \sum_{w,w'\in\Sigma^{r-1}}
g_d(L^j_w) \cdot g_d(F^{j,k}_{w,w'}) \cdot g_d(R^k_{w'}).
\]
The following lemma shows that $g_d(L^j_w)$ and $g_d(R^k_w)$ are
independent from $j,k$ and $d$.
\begin{lemma} 					\label{a_invariant_d}
For any intervals $[j,k]$, $[j',k']$ and configurations $d,d'$ such
that $\idom(d)\subseteq[j,k]$, $\idom(d')\subseteq[j',k']$ and for every
$w\in\Sigma^{r-1}$, we have
\[
	g_d(L^j_w) = g_{d'}(L^{j'}_w) \mbox{ \ \ \ and \ \ \ }
	g_d(R^k_w) = g_{d'}(R^{k'}_w).
\]
\end{lemma}

\begin{proof}
We prove only the first equation, the proof for the second one is
analogous.  Let $m = j' - j$.  We define a bijection from $L^j_w$ to
$L^{j'}_w$ which preserves the weight.  If $p = (\ldots, (w_i,i),
\ldots, (w_j,j))$ is a path in $L^j_w$, then we define its image as
$p' = (\ldots, (w_i,i+m), \ldots, (w_j,j+m))$.  This is clearly a
bijection and we also have $g_d(p) = g_{d'}(p')$ since $d_i = q$ for
$i < j$ and $d'_i = q$ for $i < j'$.
\end{proof}

We define the \emph{left} and \emph{right border vectors},
respectively $\vec l=(l_w)_{w\in \Sigma^{r-1}}$ and $\vec
r=(r_w)_{w\in \Sigma^{r-1}}$ as follows: for $w \in \Sigma^{r-1}$,
$l_w = g_d(L^j_w)$ and $r_w = g_d(R^k_w)$, where $[j,k]$ is an
arbitrary interval and $d$ an arbitrary configuration satisfying
$\idom(d)\subseteq[j,k]$.  The next lemma states that $\vec l$ and
$\vec r$ are in $\mathbb R^{\Sigma^{r-1}}$.

\begin{lemma}					\label{lem-finite}
For all $w\in\Sigma^{r-1}$, $l_w$ and $r_w$ are finite.
\end{lemma}
\begin{proof}
Suppose there is a $w$ such that $l_w =\infty$. (The case $r_w =
\infty$ is symmetric.)  We will prove that this implies the
existence of a configuration such that the associated line vector has
infinite norm, thus contradicting by Corollary~\ref{cor-path} the
hypothesis that $A$ is well-formed.

Let $w'$ be such that $r_{w'}>0$. There exists such a $w'$, since for
example $r_{q^{r-1}}\geq 1$. Let $x_1, x_2, \ldots, x_{2r-1}\in
\Sigma$ such that $w=x_1\ldots x_{r-1}$ and $w'=x_r\ldots
x_{2r-2}$. We set $w_i=x_ix_{i+1}\ldots x_{i+r-2}$ for $i=1,\ldots,r$
and $v_i=x_i x_{i+1}\ldots x_{i+r-1}$ for $i=1,\ldots,r-1$. Note that
$w_1=w$ and $w_r=w'$. For $i=1,\ldots,r-1$ let $y_i\in \Sigma$ be such
that $\transition{\delta(v_i)}{y_i}\neq 0$.  Let $j$ be an arbitrary
integer, and set $k=j+r-2$.  We define the configuration $d$ to be the
quiescent state outside the interval $[j,k]$ and $d_{j+i-1} = y_i$ for
$i=1, \ldots, r-1$.  Then in $G_\infty^d$ already the set of paths
going through the vertices $(w_1,j), \ldots , (w_r,k+1)$ has infinite
weight. Since each path has non-negative weight, $P$ has also infinite
weight.
\end{proof}

The first part of our algorithm will be the computation of the border
vectors $\vec l$ and $\vec r$.  For the second part, we reduce now our
problem to a question in linear algebra.

For every $a\in\Sigma$, let $M_a \in {\mathbb R}^{\Sigma^{r-1} \times
\Sigma^{r-1}}$ be the linear operator whose matrix is defined for all
$w,w' \in \Sigma^{r-1}$ as
\[
M_a(w',w) = \left\{ \begin{array}{ll} |\transition{\delta(xty)}{a}|^2
	& \mbox{if } w=xt, w'=ty \mbox{ for some }x,y\in\Sigma \mbox{
	and } t\in\Sigma^{r-2}, \\ 0 & \mbox{otherwise.}  \end{array}
	\right.
\]
We extend this definition to finite sequences over $\Sigma$.  If
$\epsilon$ denotes the empty word, then $M_\epsilon$ is the identity
operator.  Let $s>1$ be an integer, and $b = b_1 \ldots b_s$ be an
element of $\Sigma^s$.  We define
\[
M_{b} = M_{b_s} \cdots M_{b_1}.
\]

\begin{lemma}                                           \label{lem-vec}
Let $d$ be a configuration with $\idom(d)=[j,k]$.  Then
\[
g_d(P) =
	\langle M_{d_j \ldots d_k} \vec l|\vec r \rangle.
\]
\end{lemma}

\begin{proof}
We have
\begin{eqnarray*}
g_d(P)	&=& \sum_{w,w'\in\Sigma^{r-1}}
	g_d(L^j_w) \cdot g_d(F^{j,k}_{w,w'}) \cdot g_d(R^k_{w'})	\\
&=& \sum_{w,w'}
	l_w \cdot g_d(F^{j,k}_{w,w'}) \cdot r_{w'}			\\
&=& \sum_{w,w'}
	l_w \cdot  M_{d_j \ldots d_k}(w',w) \cdot r_{w'}	\\
&=& \sum_{w'} \left( \sum_w
	l_w \cdot  M_{d_j \ldots d_k}(w',w) \right) \cdot r_{w'}\\
&=& \sum_{w'}
	(M_{d_j \ldots d_k}\vec l) (w') \cdot r_{w'}		\\
&=& \langle M_{d_j \ldots d_k} \vec l | \vec r \rangle.
\end{eqnarray*}
\end{proof}

\noindent
Since for every $b\in\Sigma^*$, there exists a configuration $d$ whose
non-quiescent part is $b$, Corollary~\ref{cor-path} and Lemma~\ref{lem-vec}
imply the following reduction.
\begin{corollary}					\label{cor:inner}
A well-formed LQCA is unitary if and only if for every $b\in\Sigma^*$,
we have $\langle M_{b} \vec l|\vec r \rangle = 1$.
\end{corollary}

We also have the following property, which simplifies our next
reduction step.
\begin{lemma}						\label{lem-lr1}
For every well-formed LQCA $A$ we have $\langle \vec l | \vec r \rangle
=1$.
\end{lemma}
\begin{proof}
Let $d$ be the all quiescent configuration.  Then in $U_A$ the column
indexed by $d$ has the entry $1$ at row $d$ and $0$ elsewhere.  Since
the column vectors of $U_A$ are pairwise orthogonal,
Lemma~\ref{lem-iso} implies that the row indexed by $d$ has only zero
entries besides column $d$.  Therefore this row has norm $1$, and the
claim follows from Lemma~\ref{lem-weightRow} and~\ref{lem-vec}.
\end{proof}

Let $m=|\Sigma^{r-1}|$.  The border vectors can be seen as elements of
${\mathbb R}^m$, and the elements of the set ${\cal M} = \{M_a : a \in
\Sigma \}$ can also be seen as linear transformations in ${\mathbb
R}^m$.  Let us fix a few notations for the inner product space
$\mathbb R^m$.  Let $S \subseteq {\mathbb R}^m$ a finite set of
vectors.  The \emph{linear subspace} and the \emph{affine subspace}
generated by $S$, denoted here respectively by $\linHull S$ and
$\affHull S$ are defined as
\begin{eqnarray*}
\linHull S &=& \{ \lambda_1 \vec s_1  + \ldots + \lambda_t \vec s_t |\: 
	t \geq 0; 
	\vec s_1,\ldots,\vec s_t\in S; 
	\lambda_1,\ldots,\lambda_t\in\mathbb R\},			\\
\affHull S &=& \{ \lambda_1 \vec s_1 + \ldots + \lambda_t \vec s_t |\: 
	t \geq 0; 
	\vec s_1,\ldots,\vec s_t \in S; 
	\lambda_1,\ldots,\lambda_t\in\mathbb R;
	\lambda_1+\ldots+\lambda_t = 1\}.
\end{eqnarray*}
$B$ is said to be a basis of $\linHull S$ (respectively of $\affHull S$)
if $\linHull B = \linHull S$ (respectively $\affHull B = \affHull S$)
and has minimal cardinality for this property.

Let $\vec u \in {\mathbb R}^m$ be a vector and ${\cal F} \subseteq
{\mathbb R}^{m\times m}$ be a finite family of linear transformations.
We set $S + \vec u = \{\vec v+\vec u : \vec v \in S\}$, and ${\cal
F}(S) = \{f(\vec v) : \vec v\in S, f \in {\cal F}\}$.  Let $H_{\vec
u}$ denote the linear subspace whose normal vector is $\vec u$, that
is $H_{\vec u} = \{ \vec v : \langle \vec v | \vec u \rangle =0\}$.

We define by induction on $i$, for $i \geq 0$, the sets ${\cal
F}^i(S)$.  Let ${\cal F}^0(S) = S$, and ${\cal F}^{i+1}(S) = {\cal
F}^i(S) \cup {\cal F}({\cal F}^i(S))$.  We say that $S$ is
\emph{closed} for $\vec v$ under ${\cal F}$, if $\bigcup_{i=0}^\infty
{\cal F}^i(\{\vec v\}) \subseteq S$.

From Lemma~\ref{lem-lr1}, $H_{\vec r} + \vec l$ is the set of vectors
which have unit inner product with $\vec r$, that is
\[
	H_{\vec r} + \vec l = \{ \vec u : \langle \vec u |  \vec r
	\rangle =1\}.
\] 
It is an affine subspace since $H_{\vec r}+\vec l = \affHull{ H_{\vec
r}+\vec l }$.  Clearly for every $b \in \Sigma^*$, $\langle M_{b} \vec
l | \vec r \rangle = 1$ if and only if $H_{\vec r} + \vec l$ is closed
for $\vec l$ under ${\cal M}$.  Therefore by Corollary~\ref{cor:inner}
our reduction steps lead to the following theorem:
\begin{theorem}                                         \label{thm-reduc}
A well formed LQCA is unitary if and only if $H_{\vec r} + \vec l$ is
closed for $\vec l$ under ${\cal M}$.
\end{theorem}

This characterization is illustrated in figure~\ref{hyper}.

\section{Computing the border vectors}			\label{sec-border}

In this section we will give an algorithm for computing the border
vectors.  By symmetry, it will be sufficient to give it only for the
left vector.  The main tool in the computation will be the weighted
border graph.  Its underlying graph can be seen as a slight
modification of the finite version of the configuration graph.  This
graph was also used in \cite{DLS97} for checking that all the columns
of $U_A$ had unit norms.  However, there the weights were defined as
the norms of the transition state superpositions, whereas here they
will be the squared magnitudes of the amplitude of the quiescent state
in those superpositions.

	The \emph{(left) border graph} is the finite, directed,
weighted graph $G_l=(V,E,g)$.  The vertex set is $V = \Sigma^{r-1}$
and the edge set is
\[
	E = \{(xt,ty) : x,y \in \Sigma , t \in \Sigma^{r-2}\} 
\]
The weight function is defined as
\[
	g((xt,ty)) = |\transition{\delta(xty)}{q}|^2.
\]
A \emph{path} in $G_l$ is a finite, non empty sequence of at least two
vertices such that there is an edge between every two consecutive
vertices.  Observe that a single vertex alone here does not form a
path.  As usual, the weight of a path is the product of the edge
weights, and the weight of a set of paths is the sum of the individual
path weights.  The weight of the empty path set is 0.

For every $w \in \Sigma^{r-1}$, we define $P_w$ as the set of paths in
$G_l$ whose first vertex is $q^{r-1}$, whose second vertex is
different from $q^{r-1}$ and whose last vertex is $w$.

\begin{lemma}						\label{lem-l=w}
For every $w \in \Sigma^{r-1}$, we have
\[
   l_w  = \left\{ \begin{array}{ll}
                g(P_w)   & \mbox{if } w \neq q^{r-1}, \\
                g(P_w)+1 & \mbox{if } w = q^{r-1}.
                                          \end{array} \right.
\]
\end{lemma}
\begin{proof}
Let $d$ be a configuration with interval domain $[j,k]$, and let 
\[
	p_q = (\ldots, (q^{r-1},i), \ldots , (q^{r-1},j)).
\]
We set $L_w' = L_w^j - \{p_q\}$.  We will give a weight preserving
bijection from $L_w'$ to $P_w$ which maps $p$ to $p'$.  Let $p =
(\ldots, (w_i,i), \ldots , (w_j,j))$ be an element of $L_w'$, where
$w_j = w$.  Let $h \leq j$ be the greatest integer such that for every
$i \leq h$, we have $w_i = q^{r-1}$.  Then we set $p' = (q^{r-1},
w_{h+1}, \ldots , w_j)$.  This is clearly an injective mapping, and it
is also surjective since by the choice of $h$, $w_{h+1} \neq q^{r-1}$.
It is also weight preserving since the edges in $p$ until the vertex
$(w_h,h)$ have all weight 1.  Since $g_d(p_q) = 1$ the lemma follows.
\end{proof}

\begin{theorem}						\label{thm-border}
There exists an algorithm which computes the border vectors in time
$O(n^{\frac{3(r-1)}{r+1}})$.
\end{theorem}

\begin{proof}
According to Lemma~\ref{lem-l=w} it is sufficient to compute $g(P_w)$
for $w \in \Sigma^{r-1}$.  The main difficulty in this computation is
that the paths of $P_w$ are defined by a constraint which forces the
second vertex to be different from $q^{r-1}$.  The solution we propose
codes this constraint directly in the graph, which we will augment by
one vertex for this purpose.  Then we compute the total path weight
from $i$ to $j$ for all vertices $i,j$.  To do this, we will adapt a
standard algorithm which constructs the regular expression associated
to a finite state automaton.

Let $G'_l= (V',E', g')$ where $V'= V \cup \{sq^{r-2}\}$ for a letter
$s\not\in \Sigma$, 
\[
	E' = E \cup \{ (sq^{r-2}, q^{r-2}y) : y\in
		\Sigma\backslash\{q\} \},
\]
and $g'(e)=g(e)$ for all edges $e \in E$ and $g'((sq^{r-2}, q^{r-2}y))
= g((q^{r-1}, q^{r-2}y))$.  This graph is illustrated in
figure~\ref{G_bord}.  For every $w\in \Sigma^{r-1}$, let $P'_w$ be the
set of all paths in $G'_l$ from $sq^{r-2}$ to $w$.  Clearly there is a
weight-preserving bijection between $P'_w$ and $P_w$.

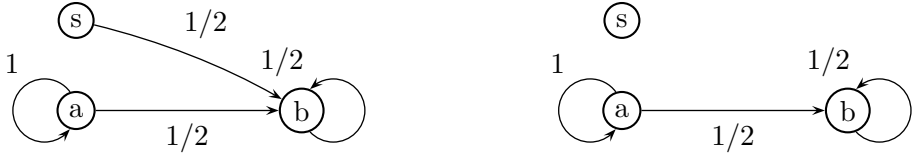
\begin{figure}[htb]
\psset{unit=6mm}
\begin{eqnarray*}
\begin{pspicture}{}(0,0)(9,4)
\cnodeput(1,1){a}{a}	
\cnodeput(6,1){b}{b}	
\cnodeput(1,3){s}{s}	
\psset{linewidth=0.5pt}
\ncarc{->}{s}{b}\Aput{1/2}
\ncline{->}{a}{b}\Bput{1/2}
\nccircle[angleA=90]{->}{a}{0.7}\Bput{1}
\nccircle[angleA=270]{->}{b}{0.7}\Bput{1/2}
\end{pspicture}
&\mbox{\hspace*{3em}}&
\begin{pspicture}{}(0,0)(9,4)
\cnodeput(1,1){a}{a}	
\cnodeput(6,1){b}{b}	
\cnodeput(1,3){s}{s}	
\psset{linewidth=0.5pt}
\ncline{->}{a}{b}\Bput{1/2}
\nccircle[angleA=90]{->}{a}{0.7}\Bput{1}
\nccircle[angleA=270]{->}{b}{0.7}\Bput{1/2}
\end{pspicture}
\end{eqnarray*}
\caption{The graphs $G'_l$ (left-hand) and $G'_r$ (right-hand),
associated to the LQCA \textsc{Qflip}.  From these graphs we can
compute $\vec l =\Choose11$ and $\vec r =\Choose10$.}
\label{G_bord}
\end{figure}

The border vectors have only finite components, nevertheless for their
computation we have to extend the non-negative real numbers with
$\infty$.  Let ${\mathbb R}^*$ be this set.  We define the following
computation rules with respect to $\infty$:
\[
	\infty + c = c + \infty = \infty,
\]
for every $c \in {\mathbb R}^*$,
\[
	\infty \cdot c = c \cdot \infty = \infty \cdot \infty = \infty,
\]
for every real number $c > 0$, 
\[
	\infty \cdot 0 = \infty \cdot 0 = 0,
\]
and $\infty^0 = 1$.  We also define $c^*$  for every $c \in {\mathbb R}^*$
as $\sum_{e'=0}^\infty c^{e'}$, that is
\[
        c^*  = \left\{ \begin{array}{ll}
                1/(1-c) & \mbox{if } 0 \leq c < 1, \\
                \infty & \mbox{otherwise.}
                                          \end{array} \right.
\]
Let $\{v_1,v_2,\ldots,v_{|V'|}\}$ be an arbitrary enumeration of the
vertices of $G'_l$.  For $1 \leq i,j \leq |V'|$ and for $0 \leq k \leq
|V'|$, we define the path sets $P_k(i,j)$ as the set of paths which
start in $v_i$, end in $v_j$, and all the other vertices in the path
have index less or equal to $k$.  Let $W_k(i,j)$ denote $g(P_k(i,j))$.
Then we claim that $W_k(i,j)$ satisfies the following recursion for $1
\leq i,j \leq |V'|$, and $1 \leq k \leq |V'|:$
\begin{eqnarray*}
        W_0(i,j)   &=& \left\{\begin{array}{ll}
                        g'((v_i,v_j))    &\mbox{if } (v_i,v_j)\in E', \\
                        0               &\mbox{otherwise,}
                                 \end{array} \right.    \\
        W_k(i,j) &=& W_{k-1}(i,j) +
                          W_{k-1}(i,k)         \cdot
                          (W_{k-1}(k,k))^*     \cdot
                          W_{k-1}(k,j).              
\end{eqnarray*}

We prove our claim by induction on $k$.  In $P_0(i,j)$ the only path
is the edge between $v_i$ and $v_j$ if this edge exists. 

Assume that this equation is true for $k-1$.  We note that for every
path of $P_k(i,j)$ there exists a unique integer $e$ such that vertex
$v_k$ appears exactly $e$ times the path. Thus we can write
\[
	P_k(i,j) = P_{k-1}(i,j) \cup \bigcup_{e=1}^\infty
		P_{k-1}(i,k) \otimes \underbrace{P_{k-1}(k,k) \otimes
	\cdots \otimes P_{k-1}(k,k)}_{e-1} \otimes P_{k-1}(k,j),
\]
where the unions are disjoint and $\otimes$ is the path composition
operator defined in section~\ref{sec-reduc}.  By induction hypothesis
we have
\[
	W_k(i,j) = W_{k-1}(i,j) +
		 \sum_{e'=0}^{\infty} \left(
				W_{k-1}(i,k) \cdot 
			(W_{k-1}(k,k))^{e'} \cdot W_{k-1}(k,j)
					\right),
\]
which concludes the induction.

This proves the correction of the following algorithm: Let
$m=|V'|=|\Sigma|^{r-1}$.  Initialize $W_0$.  For $k=1,\ldots,m$
compute $W_k$ using $W_{k-1}$.  Finally output the border vector $\vec
l$ defined by $\vec l(w)=W_m(sq^{r-2},w)$ for $w\neq q^{r-1}$ and
$\vec l(q^{r-1})=W_m(sq^{r-2},q^{r-1})+1$.  Proceed in similar fashion
for $\vec r$.  The complexity of the algorithm is
$O(|\Sigma|^{3(r-1)}) = O(n^{\frac{3(r-1)}{r+1}})$.
\end{proof}

\section{Closed affine subspace}	\label{sec-coset}

In this section we will give a polynomial algorithm for the following
problem.
\begin{quote} \samepage
\textsc{\large Closed Affine Subspace}
\begin{description}
\item[{Input:}] Two vectors $\vec l, \vec r \in \mathbb R^m$ such that
$\langle \vec l|\vec r\rangle=1$ and a set of linear transformations
${\cal M}=\{ M_a : a\in \Sigma\}$ in $\mathbb R^m$, where
$m=|\Sigma|^{r-1}$.
\item[{Question:}] 
	Is $H_{\vec r}+\vec l$ closed for $\vec l$ under ${\cal M}$,
	i.e.\ for all $b\in \Sigma^*$ do we have $M_b \vec l \in
	H_{\vec r} + \vec l$?
\end{description}
\end{quote}

We set $t = |\Sigma|$.  For the simplicity of notation, let $H=H_{\vec
r}+\vec l$ and let $E_i={\cal M}^i(\{\vec l\})$.  Since $H = \affHull
H$, we have $E_i \subseteq H$ if and only if $\affHull{E_i} \subseteq
H$.  Therefore we have to decide if $\affHull{\bigcup_{i=0}^\infty
E_i} \subseteq H$.  Dimension arguments imply the existence of a
fixpoint, a set $E_j$, such that $\affHull{E_j} =
\affHull{\bigcup_{i=0}^\infty E_i}$.  Moreover we need only to keep
track of a basis of $\affHull{E_i}$, that is a set $B_i$ of linearly
independent vectors, with $\affHull{B_i}=\affHull{E_i}$.

\begin{theorem}						\label{thm-coset}
There exists an algorithm which decides if $H$ is closed for $\vec
l$ under ${\cal M}$ in time $O(n^{\frac{3r-2}{r+1}})$.
\end{theorem}

\begin{proof} 
We claim that this is realized by the following algorithm:
\begin{quote}
	$\begin{array}{@{}l@{\;}l@{\;}l}
	B_0 &:=& \{ \vec l\} 					\\
	i   &:=&   1		
	 \end{array}$						\\
	{\bf while}  
	$\affHull{B_i} \neq 
			\affHull{ B_i \cup {\cal M}(B_i) }$ 	\\
\hspace*{3em}
		$B_{i+1}$ := a basis of
		$\affHull{B_i \cup {\cal M}(B_i) }$ 		\\
\hspace*{3em}
                $i := i+1$					\\
        $B := B_i$						\\
	{\bf if} $B \subseteq H$ accept 			\\
	{\bf else} reject
\end{quote}

At every iteration dim($\affHull{ B_i }$) increases, and
therefore the algorithm terminates in at most $m-1$ iterations.  We
prove that it is correct.

We show $\affHull{ B_i } = \affHull{ E_i }$ for every $i \geq 0$ by
induction.  The statement holds by definition for $i=0$.  Suppose
$\affHull{ E_i }=\affHull{B_i}$ for some $i$.  Since $\cal M$ contains
only linear operators, we have for any set $S$, $\affHull{{\cal M}(S)}
= \affHull{{\cal M}([S])}$.  Therefore $\affHull{{\cal M}(E_i) } =
\affHull{{\cal M}(B_i) }$ which implies $\affHull{ E_{i+1}} =
\affHull{ B_{i+1}}$.

Since $B$ is a fixpoint, that is $\affHull{ B }= \affHull{B \cup {\cal
M}(B) }$, this implies that $\affHull{ B } = \affHull{
\bigcup_{i=0}^{\infty} E_i }$ and the correctness follows.

We now turn to the analysis of the complexity.  We will build
inductively the basis so that for all $i$, $B_i \subseteq B_{i+1}$.
At the $i$-th iteration, to build $B_{i+1}$, initially we set
$B_{i+1}=B_i$.  Then we compute every vector in ${\cal M}(B_i
\backslash B_{i-1})$ and add it to $B_{i+1}$ if it is not in the
affine subspace generated by $B_{i+1}$.  At the end of the algorithm,
these steps were applied to all vectors $M \vec b$ for $M\in {\cal M}$
and $\vec b \in B$, thus at most $|{\cal M}|\cdot | B| = O(tm)$ times.
Computing $M \vec b$ takes $O(m^2)$ with standard matrix
multiplication, and checking affine independence takes also $O(m^2)$
with the algorithm described in the next section.  Thus the overall
complexity is $O(tm^3)$.  The theorem follows since $t =
n^{\frac1{r+1}}$ and $m=n^{\frac{r-1}{r+1}}$.
\end{proof}

\section{Maintaining a basis}				\label{sec-indep}

In this section we give a dynamic algorithm for the following problem.
We want to maintain a basis $B$ of a $d$-dimensional linear subspace
in $R^m$, such that the following requests for a given vector $\vec
u\in\mathbb R^m$ can be treated efficiently:
\begin{description}
\item[{Membership query:}]  
		``Is $\vec u \in \linHull{B}$?''
\item[{Add to basis:}]	    
		Replace $B$ by $B \cup\{\vec u\}$.
\end{description}

We can define the problem for affine subspaces as well.  Fortunately
the latter can easily be reduced to the former: Let $f : \mathbb R^m
\rightarrow \mathbb R^{m+1}$ be the function which maps a vector $\vec
u$ to $\vec u'$ with $u'_i = u_i$ for $i=1,\ldots,m$ and $u'_{m+1}=1$.
Then every vector $\vec v$ satisfies $\vec v \in \affHull{B}$ if and
only if $f(\vec v) \in \linHull{f(B)}$.

A solution to this problem requires a tricky data structure to encode
the basis.  The naive way would be to represent $B$ by a matrix such
that its column vectors are exactly those of $B$ and to apply the
Gaussian elimination algorithm (see for example \cite{Len90}) to check
whether $\vec u\in\langle B \rangle$.  This would require $O(md^2)$
time steps.

Transforming the matrix in an upper triangular form is the bottleneck
of this approach.  The representation we choose for $B$ will improve this
complexity.
\begin{theorem}
There is a dynamic algorithm for maintaining a basis of a
$d$-dimensional subspace in $\mathbb R^m$ which treats each request
in time $O(m(m-d))$.
\end{theorem}

\begin{proof}
We represent a non empty basis $B$ by the couple $(T,B)$, where
$T\in\mathbb R^{m\times m}$ is an orthogonal matrix and $\linHull{
T(B) } = \mathbb R ^d \times \{0\}^{m-d}$.  The empty basis is
represented by $(I, \emptyset)$, where $I$ is the identity matrix.
                                                    
Since $T(\linHull{ B }) = \linHull{ T(B) }$, $\vec u \in \linHull{ B
}$ if and only if $T\vec u\in\mathbb R^d \times \{0\}^{m-d}$.  Thus
verifying $\vec u\in \linHull{ B }$ is reduced to checking if the last
$m-d$ components of $T\vec u$ are all $0$ which can be done in time
$O(m(m-d))$.

Suppose $\vec u \not\in \linHull{ B }$.  We will show that there is a
orthogonal matrix $M$ affecting only components from $d+1$ to $m$
which satisfies $(MT\vec u)_{d+1} \neq 0$ and $(MT \vec u)_i=0$ for
$i=d+2, \ldots, m$.  Thus $\linHull{ MT(B) } = \linHull{ T(B) }$ and
$\linHull{ M T (B \cup \{ \vec u \}) } = \mathbb R ^{d+1} \times \{0\}
^{m-(d+1)}$.  Therefore $(MT, B \cup \{ \vec u \})$ represents
$B\cup\{\vec u\}$.

We define $M$ as the composition of two operators $M_1$ and $M_2$ we
describe now.  By hypothesis $\vec u \not\in \linHull{ B }$, therefore
there exists an index $k\in\{d+1,\ldots,m\}$ such that $(T\vec
u)_k\neq 0$.  Define $M_1$ to be the permutation matrix which
exchanges $k$ and $d+1$.

Let $\vec u' = M_1 T\vec u$.  Note that $\vec u'_{d+1} \neq 0$.  Then
for an arbitrary vector $\vec v$ we define $(M_2\vec v)_i=\vec v_i$
for $i=1, \ldots, d+1$ and $(M_2 \vec v)_i = \vec v_i - \vec v_{d+1}
\vec u'_i / \vec u'_{d+1} $ for $i=d+2,\ldots,m$.  Clearly $M_1$ and
$M_2$ are orthogonal linear operators, and since
\[
        M_2 \vec u' = (u'_1, u'_2, \ldots, u'_{d+1}, 0, \ldots, 0)
\]     
$M$ satisfies the required property.  We can compute $M_2 M_1 T$ in
time $O(m(m-d))$, which concludes the proof.
\end{proof}

\section{Putting all together}				\label{sec-together}

We are now able to prove Theorem~\ref{thm-main}, that is to give an
algorithm to decide if a given LQCA is unitary.  By
Theorem~\ref{thm-reduc} to solve this problem we have to compute the
associated border vectors and decide the corresponding \textsc{Closed
Affine Subspace} problem.  According to Theorem~\ref{thm-border}
the border vectors can be computed in time
$O(n^{\frac{3(r-1)}{r+1}})$, and due to Theorem~\ref{thm-coset} the
last problem can be solved in time $O(n^{\frac{3r-2}{r+1}})$, which
concludes the proof.

\section{Conclusion}					\label{sec-concl}

A not necessarily simple LQCA can be transformed into a simple one
with the same time evolution operator.  Let the original neighborhood
be $N=(a_1, \ldots, a_r)$.  The size of the new neighborhood will be
$s= a_r -a_1 +1$.  If we define the \emph{expansion factor} of an LQCA
as $e = (s+1)/(r+1)$ then the algorithm works in the general case in
time $O(n^{e\frac{3r-1}{r+1}}) = O(n^{3e})$.

In the case of space-periodic configurations van~Dam \cite{Dam96} has
shown that LQCAs can be efficiently simulated by QTMs.
Watrous~\cite{Wat95} gave an equivalent result for partitioned LQCAs.
This question remains still open for the model of this paper.

\section*{Acknowledgments}

We are thankful to Laurent Rosaz, Huong L\^eThanh and Umesh Vazirani
for several helpful conversations.  We wish also to thank the
anonymous referees for helpful comments and in particular for
simplifying section~\ref{sec-coset}.


\end{document}